\documentclass[twocolumn,secnumarabic, showpacs, preprintnumbers, amssymb, floatfix, prb]{revtex4}
\usepackage{epsf}
\usepackage{bm}

\renewcommand{\Im}{\mathop{\rm Im}\nolimits}
\renewcommand{\Re}{\mathop{\rm Re}\nolimits}

\begin{document}

\author{V. A. Margulis and M. A. Pyataev}
\email{pyataevma@math.mrsu.ru}

\affiliation{Mordovian State University, Saransk, Russia}.

\title{Electron transport on a cylindrical surface
with one-dimensional leads}

\date{\today}

\begin{abstract}
A nanodevice consisting of a conductive cylinder
in an axial magnetic field with one-dimensional wires
attached to its lateral surface is considered.
An explicit form for transmission and reflection coefficients of
the system as a function of electron energy is found
from the first principles.
The form and the position of transmission
resonances and zeros are studied.
It is found that, in the case of one wire
being attached to the cylinder, reflection peaks occur at
energies coinciding with the discrete part of the
electronic spectrum of the
cylinder. These peaks are split in a magnetic field.
In the case of two wires
the asymmetric Fano-type resonances are detected
in the transmission between the wires
for integer and half-integer values of the magnetic flux.
The collapse of the resonances appears
for certain position of contacts.
Magnetic field splits transmission peaks
and leads to spin polarization of transmitted electrons.
\end{abstract}


\maketitle

\section{Introduction}

Electron transport in curved two-dimensional nanostructures
attracts considerable attention in the last decade. Transport
properties of the electron gas on spherical
\cite{FLP95,Kis97,BGM02,GMP03,MP04} and cylindrical
\cite{CMR98,MRD98} nanosurfaces are intensively studied in the
last few years. Those systems are of particular interest due to
recent intensive experimental investigations of the coherent
transport in individual carbon nanotubes
\cite{KYT01,KKL03,BS04,GKE04,AKR04} and rolled GaAs/AlGaAs
heterostructures \cite{Pri00,VPY04}. A number of theoretical works
\cite{Nar99, AG99, CI99, SKT00, FCR01, GFC02, Ury04, KPP05} has been
focused on the electron transport in carbon nanotubes (see
Ref.~\onlinecite{Ando00} for review). It should be mentioned that
nanotube-based multiterminal nanodevices attract recently more
and more attention since they are proposed as promising units for
future low-power high-speed electronics. A lot of works
\cite{EK00,PPG00,CX02,SB02,Xu02,CYP02,CX03} are devoted to the
investigation of the electron transport in various interesting
multiterminal nanodevices.

The conductance is usually measured for two basic geometries of contacts.
Most of theoretical studies
are focused on the case of end-contacted nanotubes.
In this geometry a strong interaction
between metal and carbon atoms is realized
resulting in low contact resistance.
However in the last few years much attention
is devoted to side-contacted nanotubes \cite{CCC03}.
In this case the leads are attached to the lateral
surface of the tube.
The interest to these structures is stipulated by recent
experiments on scanned probe microscopy
of electronic transport in the nanotubes \cite{BFP00,YPR04}.
The tip of the atomic force microscope can play the
role of the side-contacted lead.
Another interesting system with laterally attached leads
is branched 'nanotree' reported in Ref.~\onlinecite{DDL04}.
We mention that the side-contacted geometry is also realized
in crossed carbon nanotubes \cite{KKL03}.
It is evident that the lateral disposition of the contacts
can significantly affect the transport.
In particular, the resonant transport regime
is expected in this case.

Recent experiments on
the transport in carbon nanotubes \cite{KKL03,BS04}
have reported  the presence of
asymmetric Fano resonances
in the dependence of conductance on the Fermi energy.
Being a characteristic manifestation of wave phenomena
in scattering experiments
resonances have received considerable attention in recent
electron transport investigations.
A number of papers
\cite{KS99, KSJ99, KS99E, KSR02}
is devoted to the study of Fano resonances in the transport
through quasi-one-dimensional
channels with impurities.
It is shown in Refs.~\onlinecite{GMP03, MP04}  that the same resonances occur
in the conductance through a quantum nanosphere and a quantum nanotorus.
Similar phenomena could be expected in the electron
transport through the quantum cylinder
but our analysis shows that the form
of resonances differs from the Fano line shape.

The purpose of the present
paper is an investigation of the
electron transport through a multiterminal nanodevice consisting
of a conductive cylindrical surface ${\rm  C}$ with
one-dimensional wires attached to it. The cylinder is placed in an
axial magnetic field $B$ and the wires are attached to its lateral
surface. The number of wires we denote by $N$. We consider in
detail the case of one and two wires attached to the cylinder. The
points of contacts on the cylinder we denote by ${\bf
q}_j=(z_j,\varphi_j)$, where $z_j$ and $\varphi_j$ are cylindrical
coordinates and  $j=1,\ldots, N$ is the number of the contact.

In our model, the electron on the cylinder
is able to go away from the contact region to infinity
and never returns back. We stress, that the model
is valid for a realistic finite-size cylinder if its
bases are immersed into absorbing electron reservoirs as shown
in Fig.~\ref{f-scheme}.

\section{Hamiltonian and transmission coefficient}

\begin{figure}[!h]
\epsfclipon
\epsfxsize=64mm
\begin{center}
\epsffile{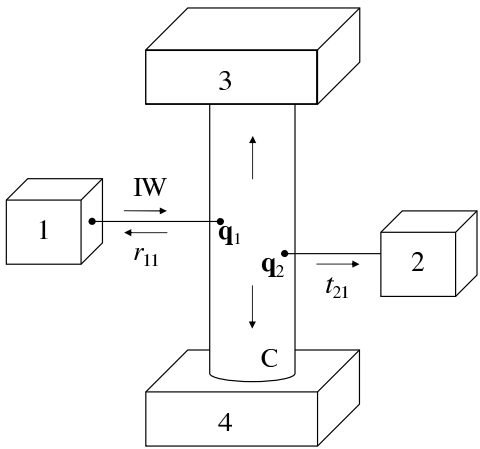}
\end{center}
\caption{\label{f-scheme}
Scheme of the device in the case
of two wires attached to the cylinder.
An incident wave (IW) originating from reservoir 1
is reflected back with amplitude $r_{11}$
and scattered to reservoir 2 with amplitude $t_{21}$.
Reservoirs 3 and 4 absorb the electron waves
going away from the contact region.}
\end{figure}

In the model, the wires are taken to be one-dimensional
and represented by semiaxes
${\mathbb R}^+_j=\{x: x\geq 0\}$ ($j=1 \ldots N$).
They are connected to the cylinder by gluing the point
$x=0$ from ${\mathbb R}^+_j$ to the point ${\bf q}_j$ from ${\rm  C}$.
We suppose ${\bf q}_i\neq{\bf q}_j$ for $i\neq j$.
The scheme of the device is shown in Fig.~\ref{f-scheme}.

If spin-orbital interaction is absent then spin orientation conserves
and transmission coefficients
$T^\uparrow(E)$ and $T^\downarrow(E)$
for electrons polarized in the direction of the magnetic field and
in the opposite direction
may be expressed in terms of the transmission
coefficient $T(E)$ for spin-free scattering
\begin{displaymath}
T^\uparrow(E)=T(E-g \mu_B B/2),\quad
T^\downarrow(E)=T(E+g \mu_B B/2),
\end{displaymath}
where $g$ is electron g-factor and $\mu_B$ is the
Bohr magneton. Similar relations are valid for reflection
coefficients. Further, we will deal with the spin-free problem and
use spin indices only where it is necessary.

A wavefunction $\psi$ of the electron in the device
consists of $N+1$ parts:
\begin{equation}
\psi=
\left(\begin{array}{c}
\psi_{\rm  C}\\
\psi_1\\
\dots\\
\psi_N
\end{array}
\right),
\end{equation}
where
$\psi_{\rm  C}$ is a function on ${\rm  C}$
and $\psi_j$ $(j=1,\ldots,N)$ are functions on ${\mathbb R}^+_j$.

To obtain the Hamiltonian $H$ of the whole system
we use an approach  based on the operator extension theory
\cite{ES87,BG03}.
This method has been already used in Refs.~\onlinecite{BGM02,GMP03,MP04}
for the investigation of the electron transport through
the nanosphere and the nanotorus.

The Hamiltonian $H$ of the whole system is a point perturbation
of the operator
\begin{equation}
H_0=H_{\rm C}\oplus H_1\oplus\ldots\oplus H_N,
\end{equation}
where $H_{\rm  C}$  is an electron Hamiltonian on the cylinder
and $H_j$ are Hamiltonians in the wires ${\mathbb R}^+_j$.

Using cylindrical coordinates, we can represent
the Hamiltonian $H_{\rm  C}$
in the form
\begin{equation}
H_{\rm  C}=\frac{p_z^2}{2 m_{\rm c}}+
\frac{\hbar^2}{2 m_{\rm c} r^2}
\left(i{\partial\over\partial\varphi}-{\Phi\over\Phi_0} \right)^2,
\end{equation}
where $p_z$ is the $z$-component of the momentum,
$r$ is the radius of the cylinder,
$ m_{\rm c}$ is the electron effective mass on the cylinder,
$\Phi=\pi r^2 B$ is the magnetic flux,
and $\Phi_0=2\pi\hbar c / |e|$ is the magnetic flux quantum.
It is convenient to represent the Hamiltonian
$H_{\rm  C}$ in the form
$H_{\rm  C}=H_z+H_\varphi$
where $H_z={p_z^2}/{2 m_{\rm c}}$ and
$H_\varphi=\frac{\hbar^2}{2 m_{\rm c} r^2}
(i{\partial\over\partial\varphi}-\Phi/\Phi_0)^2$.
We will need below
the eigenvalues
\begin{equation}
E_m=\frac{\hbar^2}{2 m_{\rm c} r^2}\left(m+{\Phi\over \Phi_0}\right)^2
\end{equation}
and the eigenfunctions
\begin{equation}
\Psi_m(\varphi)=(2\pi r )^{-1/2}\exp (i m \varphi)
\end{equation}
of the operator $H_\varphi$.

Electron motion in each wire ${\mathbb R}^+_j$
is described by the Hamiltonian  $H_j=p_x^2/2{m_{\rm w}}$,
where $p_x$ is the momentum operator
and $m_{\rm w}$ is the effective mass for the electron in the wires.

To define the Hamiltonian $H$  we use boundary conditions at points of gluing.
The role of boundary values for the wavefunctions $\psi_j$  is played,
as usual, by $\psi_j(0)$  and $\psi_j'(0)$.
The zero-range potential theory shows that to obtain a non-trivial
Hamiltonian for the whole system we must consider functions $\psi_{\rm  C}$
with a logarithmic singularity at the points of gluing ${\bf q}_j$
\begin{equation}
            \label{asymp}
\psi_{\rm  C}({\bf x})=
-u_j{{m_{\rm c}} \over \pi \hbar^2}\ln\rho({\bf x},{\bf q}_j)+v_j +o(1),
\; \text{as}\; {\bf x} \to {\bf q}_j.
\end{equation}
Here $u_j$ and $v_j$ are complex coefficients and $\rho({\bf x},{\bf q}_j)$
is the geodesic distance on the cylinder between the points
${\bf x}$ and ${\bf q}_j$.
It is known, that the most general self-adjoint boundary
conditions are defined by some linear relations between
$\psi_j(0)$,  $\psi_j'(0)$,
and the coefficients $u_j$ and $v_j$.
Following Ref.~\onlinecite{GMP03}, we will write these conditions in the form
\begin{equation}
                       \label{bound}
\cases
{v_j=\sum\limits_{k=1}^{N}\left[ B_{jk}u_k-
\frac{\hbar^2}{2 m_{\rm w}}A_{jk }\psi'_k(0)\right],
& ${j=1\ldots N,}$ \cr
\psi_j(0)=\sum\limits_{k=1}^{N}\left[ A_{kj}^* u_k-
\frac{\hbar^2}{2 m_{\rm w}}C_{jk}\psi'_k(0)\right].  \cr}
\end{equation}
Here complex parameters
$A_{jk}$, $B_{jk}$, and $C_{jk}$ are the elements of $N\times N$
matrices.
The matrices $B$ and $C$ have to be Hermitian because the Hamiltonian $H$
is a self-adjoint operator \cite{BG03}.
To avoid a non-local tunnelling coupling between
different contact points we will restrict ourselves
to the case of diagonal matrices $A$, $B$, and $C$ only.
According to the zero-range potential theory
diagonal elements of the matrix $B$ determine the strength
of point perturbations of the Hamiltonian $H_{\rm  C}$ at the points
${\bf q}_j$ on the cylinder.
These elements may be expressed in terms
of scattering lengths $\lambda^B_j$
on the corresponding point perturbations:
$B_{jj}={m_{\rm c}}\ln(\lambda^B_j)/\pi\hbar^2$.
Similarly elements $C_{jj}$ describe the strength
of point perturbations at the point $x=0$ in the wires
and may be expressed in terms of scattering lengths  $\lambda^C_j$
by the relation  $C_{jj}= -{m_{\rm w}}\lambda^C_j/2\hbar^2$.
For convenience, we represent parameters $A_{jj}$
in the form
$A_{jj}={m_{\rm w}}\sqrt{\lambda^A_j}e^{ i \phi_j}/\hbar^2$,
where $\lambda^A_j$ has the dimension of length
and $\phi_j$ is the argument of the complex number $A_{jj}$.
We mention that the effect of the scattering
lengths $\lambda^A_j$, $\lambda^B_j$, and $\lambda^C_j$
on the electron transport has been discussed in Ref.~\onlinecite{GMP03}.
In the present paper we concentrate our attention on phenomena
which are independent of contact parameters.

To obtain transmission and reflection coefficients
of the system one needs a solution of the Schr\"odinger equation
for the Hamiltonian $H$.
The function $\psi_1$ in this solution is a superposition
of incident and reflected waves while
other functions $\psi_j$ $(j=2,\ldots N)$ represent scattered waves.
The wavefunction $\psi_{\rm  C}$ may be expressed \cite{GMP03} in terms
of the Green function $G({\bf x}, {\bf x}'; E)$
for the Hamiltonian $H_{\rm  C}$.

\begin{equation}
        \label{solution}
\left\{
\begin{array}{l}
\psi_{\rm C}({\bf x})=
\sum\limits_{j=1}^N\xi_j(E) G({\bf x},{\bf q}_j;E),\\
\psi_j(x)= \delta_{j1}e^{- i  kx}+S_{j}(E)e^{ i  kx},
\quad j=1,\ldots,N.
\end{array}\right.
\end{equation}
Here $k=\sqrt{2m_{\rm w} E/\hbar^2}$ is the electron wave vector in wires,
$\xi_j(E)$ are complex factors,
and $S_j(E)$ is the amplitude
of the outgoing wave in the wire ${\mathbb R}^+_j$.

It is well known that the Green function $G({\bf x}, {\bf x}'; E)$
may be represented in the form
\begin{equation}
\label{G}
G({\bf x}, {\bf x}'; E)=
\sum\limits_{m=-\infty}^{\infty}G_z(z,z';E-E_m)
\Psi_m(\varphi)\Psi^*_m(\varphi'),
\end{equation}
where ${\bf x}=(z, \varphi)$, ${\bf x}'=(z', \varphi')$, and
\begin{equation}
\label{Gz}
G_z(z, z'; E)={i m_{\rm c}\over\hbar^2 k}e^{ik|z-z'|}
\end{equation}
is the Green function of the operator $H_z$.
Substituting (\ref{Gz}) into (\ref{G}), we get the following equation for
$G({\bf x},{\bf x}'; E)$:
\begin{equation}
\label{G1}
G({\bf x}, {\bf x}'; E)={i m_{\rm c}\over 2 \pi \hbar^2}
\sum\limits_{m=-\infty}^{\infty}
\frac{e^{ik_m|z-z'|+im(\varphi-\varphi')}}{k_m  r },
\end{equation}
where $\hbar^2 k_m^2=2 m_{\rm c}(E-E_m)$, $\Re k_m>0$ for $E>E_m$
and $\Im k_m>0$  for $E<E_m$.

Considering the asymptotics (\ref{asymp}) of
$\psi_{\rm  C}({\bf x})$  from (\ref{solution}) near the point ${\bf q}_j$,
we have
\begin{equation}
u_j=\xi_j,\qquad v_j=\sum\limits_{i=1}^{N} Q_{ji}(E)\xi_i.
\end{equation}
Here $Q_{ij}(E)$ is the Krein's ${\cal Q}$-function, that is $N\times N$
matrix with elements
\begin{displaymath}
Q_{ij}=\cases{
G({\bf q}_i ,{\bf q}_j ;E), &  $i\ne j$;\cr
\noalign{\medskip}
\displaystyle\lim\limits_{{\bf x} \to {\bf q}_j}
\left[
G({\bf q}_j ,{\bf x} ;E)
+\frac{m_{\rm c}}{\pi\hbar^2} \ln \rho({\bf q}_j,{\bf x})
\right]\,, &  $i=j$.\cr}
\end{displaymath}

Using the elementary relation
$$
\sum\limits_{n=1}^{\infty}\frac{\exp(-nx)}{n}=-\ln(1-e^{-x}),
$$
we can subtract the logarithmic singularity from
$G({\bf x},{\bf x}'; E)$ and get the following form
for diagonal elements of ${\cal Q}$-matrix
\begin{equation}
                \label{Qjj}
Q_{jj}=\frac{m_{\rm c}}{2\pi\hbar^2}
\left[\frac{i}{k_0 r }+
\sum\limits_{m=1}^{\infty}
\left(\frac{i}{k_{m} r }+\frac{i}{k_{-m} r }-\frac{2}{m}
\right)+2\ln r
\right],
\end{equation}
The similar method has been used in Ref.~\onlinecite{EGS96}
for calculating the ${\cal Q}$-function for electron Hamiltonian on a strip.
It should be mentioned that Eq.~(\ref{Qjj})
gives the ${\cal Q}$-function for the
free particle on a plain \cite{GMC95} in the case of
$B=0$ and $r\to \infty$.

Let us consider the asymptotics of $\psi_{\rm  C}({\bf x})$
at $z\to \pm\infty$.
As it follows from (\ref{solution}) and (\ref{G1})
the wavefunction $\psi_{\rm  C}({\bf x})$ is a superposition of
propagating modes
$$
{\widetilde\Psi_m^\pm}(\varphi,z)=\Psi_m(\varphi)\exp({\pm ik_m z}).
$$
The highest and lowest numbers of the occupied modes
we denote by $M^\pm=[\pm kr-\Phi/\Phi_0]$, where
$[x]$ means the integer part of $x$.
Using equations (\ref{solution}) and (\ref{G1}), we obtain
\begin{equation}
\psi_{\rm  C}({\bf x})\simeq
\sum\limits_{m=M^-}^{M^+}
t_m^\pm {\widetilde\Psi_m^\pm}(\varphi,z),
\end{equation}
where the sign 'plus' corresponds to $z\to +\infty$
and the sign 'minus' should be taken for $z\to -\infty$.
Here $t_m^\pm$ is the partial transmission amplitude to the mode
${\widetilde\Psi_m^\pm}(\varphi,z)$.
As follows from (\ref{solution}), the amplitude is
given by
\begin{equation}
                \label{tpm}
t_m^\pm={i\over \sqrt {2 \pi r }k_m }
\sum\limits_{j=1}^N \widetilde\xi_j e^{\mp ik_m z_j-im\varphi_j},
\end{equation}
where $\widetilde\xi_j=\xi_j m_{\rm w}/\hbar^2$.

Denote the reflection coefficient to the wire
${\mathbb R}^+_j$ by $R_{11}=|S_{1}|^2$
and the transmission coefficient by $T_{j1}=|S_{j}|^2$.
The partial transmission coefficient $T_m^\pm$ to the propagating mode
$\widetilde \Psi^\pm_m$ is defined by
$ T_m^\pm=(k_m/k)|t_m^\pm|^2 $.
We stress that the relation
\begin{equation}
R_{11}+\sum\limits_{j=2}^N T_{j1}+
\sum\limits_{m=M^-}^{M^+}(T_m^+ +T_m^-)=1
\end{equation}
is valid for an arbitrary energy $E$ that is the manifestation of
the current conservation law for our system.

Substituting (\ref{solution}) into (\ref{bound}),
we get a system of $2N$ linear equations
for $S_{j}$  and $\xi_j$
\begin{equation}
                       \label{system}
\left\{\begin{array}{l}
\sum\limits_{l=1}^{N}
Q_{jl}\xi_l=B_{jj}\xi_j-
\frac{ik\hbar^2 A_{jj}}{2 m_{\rm w}}(S_{j}-\delta_{j1})\\
S_{j}+\delta_{j1}=A^*_{jj}\xi_j-
\frac{ik\hbar^2 C_{jj}}{2 m_{\rm w}}(S_{j}-\delta_{j1})
\end{array}
\right.
\end{equation}

For convenience, we  introduce the  dimensionless elements
of ${\cal Q}$-matrix
\[
{\widetilde Q}_{ij}(E)=(\hbar^2/m_{\rm w})(Q_{ij}(E)-B_{ij}).
\]
System (\ref{system}) may be decomposed to a system of $N$ equations
for $\xi_{l}$
\begin{equation}
                       \label{system_xi}
\sum\limits_{l=1}^{N}
\left[\widetilde{Q}_{jl}-
\frac{2k\lambda^A_j\delta_{jl}}{k\lambda^C_j+4 i }\right]
\widetilde{\xi}_l=-
\frac{4k\sqrt{\lambda^A_1}e^{ i \phi_1}}
{k\lambda^C_1+4 i }
\delta_{j1}
\end{equation}
and a similar system for $S_{l}$
\begin{eqnarray}
                       \label{system_s}
\sum\limits_{l=1}^{N}
&&\frac{\sqrt{\lambda^A_1}(k\lambda^C_l+4 i )e^{ i \phi_l}}
{\sqrt{\lambda^A_l}(k\lambda^C_1-4 i )e^{ i \phi_1}}
\left[\widetilde{Q}_{jl}-
\frac{2k\lambda^A_j\delta_{jl}}{k\lambda^C_j+4 i }\right]S_l=
\nonumber\\
&&\widetilde{Q}_{j1}-\frac{2k\lambda^A_1\delta_{j1}}{k\lambda^C_1-4 i }.
\end{eqnarray}

The solutions of systems (\ref{system_xi}) and (\ref{system_s})
may be represented in the form
\begin{equation}
                \label{solut}
\xi_n=\frac{\Delta_{\xi n}}{\Delta}, \qquad
S_n=\frac{\Delta_{S n}}{\Delta}
\frac{\sqrt{\lambda^A_n}(k\lambda^C_1-4 i )e^{ i \phi_1}}
{\sqrt{\lambda^A_1}(k\lambda^C_n+4 i )e^{ i \phi_n}},
\end{equation}
where
\begin{equation}
\Delta=\det\left[\widetilde{Q}_{jl}-
\frac{2k\lambda^A_j\delta_{jl}}{k\lambda^C_j+4 i }\right],
\end{equation}

\begin{eqnarray}
\Delta_{\xi n}=&&\det\left[
\left(\widetilde{Q}_{jl}-\frac{2k\lambda^A_j\delta_{jl}}{k\lambda^C_j+4 i }
\right)\left(1-\delta_{nl}\right)\right.+
\nonumber\\
&&\left.\frac{4k\sqrt{\lambda^A_1}e^{ i \phi_1}}
{k\lambda^C_1+4 i }
\delta_{j1}\delta_{nl}
\right],
\end{eqnarray}
and
\begin{eqnarray}
\Delta_{S n}=&&\det\left[
\left(\widetilde{Q}_{jl}-\frac{2k\lambda^A_j\delta_{jl}}{k\lambda^C_j+4 i }
\right)\left(1-\delta_{nl}\right)\right.+
\nonumber\\
&&\left.\left(\widetilde{Q}_{j1}-\frac{2k\lambda^A_1 \delta_{j1}}
{k\lambda^C_1-4 i }\right)
\delta_{nl}
\right].
\end{eqnarray}

\section{Results and discussion}

Let us consider in detail the case of one wire attached to the cylinder.
Using equations (\ref{tpm}) and (\ref{solut}), we obtain
\begin{equation}
                \label{Tm}
T_m^\pm=\frac{8 k\lambda^A_1}
{\pi r  k_m|2k\lambda^A_1-(k\lambda^C_1+4 i )\widetilde{Q}_{11}|^2}.
\end{equation}
Note that  $T^+_m=T^-_m$ for any energy $E$,
i. e. the scattering is isotropic in the $z$-direction.
Reflection amplitude $r_{11}\equiv S_1$ may be obtained from
Eq.~(\ref{system_s})
\begin{equation}
                \label{r11}
r_{11}=\frac{(k\lambda^C_1-4 i )\widetilde{Q}_{11}-2k\lambda^A_1}
{(k\lambda^C_1+4 i )\widetilde{Q}_{11}-2k\lambda^A_1}.
\end{equation}
\begin{figure}[!h]
{\centering
\epsfclipon
\epsfxsize=84mm
\epsffile{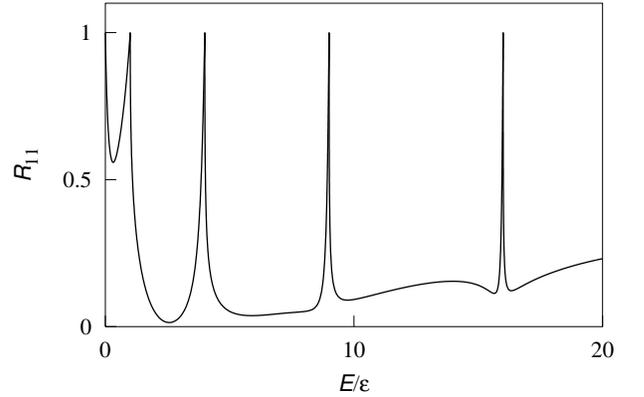}
\caption{\label{f-r11}
Reflection coefficient as a function of the electron energy
in the case of one wire attached to the cylinder
at $\lambda^A_j=\lambda^B_j=\lambda^C_j=0.4 r $, $B=0$.}}
\end{figure}
\begin{figure}[!h]
{\centering
\epsfclipon
\epsfxsize=84mm
\epsffile{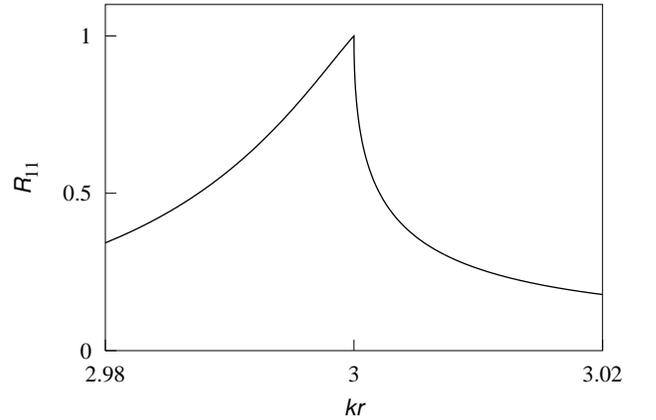}
\caption{\label{f-break}
Reflection coefficient as a function of the dimensionless parameter
$k r $. All parameters are the same as in Fig.~\ref{f-r11}.}}
\end{figure}

The reflection coefficient $R_{11}=|r_{11}|^2$ as a function of
the electron energy $E$
is represented in Fig.~\ref{f-r11}.
Hereafter we use $\varepsilon=\hbar^2/(2m_{\rm c} r^2)$
for the unit of energy and suppose $m_{\rm w}=m_{\rm c}$.
The figure shows that the dependence
of reflection coefficient on $E$ contains a series of sharp peaks
at the points $E_m$.
To study the behavior of the reflection coefficient
in a vicinity of the eigenvalues $E_m$
we consider the asymptotics of $\widetilde{Q}_{jl}(E)$ near these points
\begin{equation}
            \label{asympQ}
\widetilde{Q}_{jl}(E)\simeq \frac{\alpha_{jl}^{(m)}}{k_m}
+\beta_{jl}^{(m)}+O(k_m),\quad
\text{as}\; k_m \to 0,
\end{equation}
where
\begin{equation}
                \label{alpha}
\alpha_{jl}^{(m)}= i  \sum\limits_{m'}
\Psi_{m'}(\varphi_j)\Psi_{m'}^*(\varphi_l)
\end{equation}
and $m'$ are indices for which $E_{m'}=E_{m}$.
If the magnetic flux $\Phi/\Phi_0$ is
integer or half-integer then
the eigenvalues $E_m$ are double-degenerated and the
sum in Eq.~(\ref{alpha}) contains
two terms; otherwise it contains one term only.

As follows from Eq.~(\ref{asympQ}), the denominator in Eq.~(\ref{Tm})
has a root singularity at $E=E_m$ while the numerator remains finite.
Therefore all transmission coefficients $T^\pm_{m'}$ vanish and
the reflection coefficient $R_{11}$ reaches a unity.
The reflection coefficient
has a kink in a vicinity of each point $E_m$
stipulated by the root singularity of
the Green function on the cylinder.
Using equations (\ref{r11}) and (\ref{asympQ}), we
obtain the following asymptotics for $R_{11}(k)$ near
$\kappa_m=\sqrt{2 m_{\rm w} E_m}/\hbar$.
\begin{displaymath}
R_{11}(k)=
\cases{1-a_1 (\kappa_m-k)+o(k_m^2), &  as $k\to\kappa_m-0$\cr
1-a_2\sqrt{k^2-\kappa_m^2}+o(k_m), & as $k\to\kappa_m+0$,}
\end{displaymath}
where $a_1$ and $a_2$ are positive numbers.
The form of the reflection coefficient in a vicinity of
the point $E_3$ is shown in Fig.~\ref{f-break}.

\begin{figure}[!h]
{\centering
\epsfclipon
\epsfxsize=84mm
\epsffile{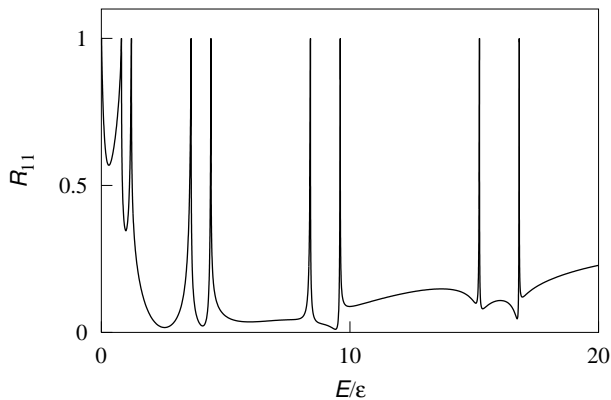}
\caption{\label{f-r11-2}
Reflection coefficient
at $\Phi/\Phi_0=0.1$.
Other parameters are the same as in Fig.~\ref{f-r11}.
}}
\end{figure}

The magnetic field splits double degenerated energy levels of the operator
$H_\varphi$ and peaks on the plot $R_{11}(k r )$ transform into
doublets (Fig.~\ref{f-r11-2}).
If the magnetic flux $\Phi/\Phi_0$ is half-integer
then the levels $E_m$ are double-degenerated and
the peaks are singlet
as for the case of integer flux,
although their positions are shifted.

\begin{figure}[!h]
{\centering
\epsfclipon
\epsfxsize=84mm
\epsffile{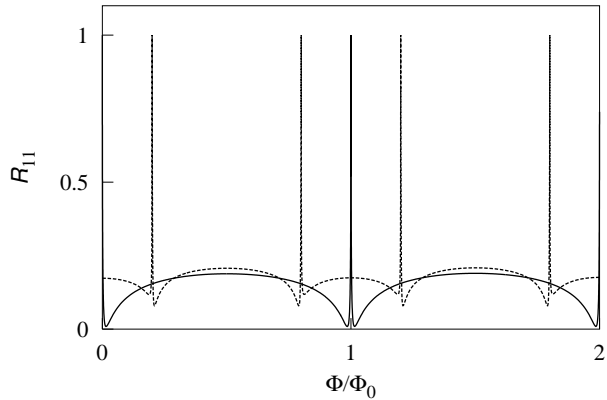}
\caption{\label{f-r11H}
Reflection coefficient as a function of the magnetic field.
Solid line: $k=4/ r $; dash line:  $k=4.2/ r $.}}
\end{figure}

The reflection coefficient as a function of the magnetic field is shown in
Fig.~\ref{f-r11H}.
Peaks on the plot correspond to the coincidence of the electron energy
with the values $E_m$.
Note that the function $R_{11}(\Phi)$ is periodic with a period $\Phi_0$
that causes the Aharonov--Bohm oscillations in the transport.
If the value of $k r $ is integer or half-integer then
there is only one peak of $R_{11}(\Phi)$ on the period,
otherwise there are two peaks on each cycle.

\begin{figure}[!h]
{\centering
\epsfclipon
\epsfxsize=84mm
\epsffile{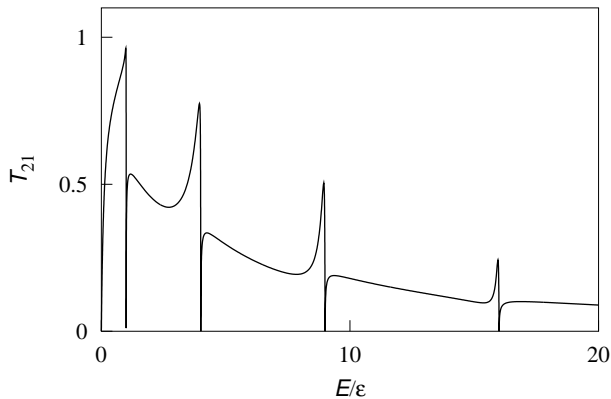}
\caption{\label{f-T21}
Transmission coefficient $T_{21}$ as a function of
the electron energy
at $\varphi_1-\varphi_2=0.08 \pi$, $z_1-z_2=0.2 r$,
and $\lambda^A_j=\lambda^B_j=\lambda^C_j=0.4 r $.}}
\end{figure}

Let us turn to the case of two wires attached to the cylinder.
Using Eq.~(\ref{solut}), we obtain
\begin{equation}
        \label{trans}
t_{21}=\frac{16ik\sqrt{\lambda^A_1\lambda^A_2}
e^{ i (\phi_1-\phi_2)}{\widetilde Q}_{21}}
{(k\lambda^C_1+4 i )(k\lambda^C_2+4 i )\Delta},
\end{equation}
where
\begin{eqnarray}
\Delta&=&\det \widetilde{Q} -\frac{2k\lambda^A_{1}}{k\lambda^C_1+4 i }\widetilde{Q}_{22}
-\frac{2k\lambda^A_{2}}{k\lambda^C_2+4 i }\widetilde{Q}_{11}
+\nonumber\\
&&\frac{4k^2\lambda^A_{1}\lambda^A_{2}}
{(k\lambda^C_1+4 i )(k\lambda^C_2+4 i )},
\end{eqnarray}
The transmissions coefficient $T_{21}=|t_{21}|^2$
as a function of
the electron energy
is shown in Fig.~\ref{f-T21}.
The figure corresponds to the case when the contacts are
placed on different generatrices ($\varphi_1\neq\varphi_2$)
and shifted along the axis of the cylinder ($z_1\neq z_2$)
in the zero magnetic field.
One can see a series of zeros
at $E=E_m$.
Transmission coefficient has a peak in the neighborhood of each zero.
The behavior of the transmission amplitude
in a vicinity of the eigenvalues $E_m$
depends strongly on contact position and applied magnetic field.
If the points ${\mathbf q}_j$ are placed on the cylinder in a random manner
then the transmission coefficient vanishes
at the double-degenerated values $E_m$.
The denominator in Eq.~(\ref{trans}) has a pole
at $E=E_m$ while the numerator has a root singularity only.
Hence, the transmission coefficient vanishes
in these points.

Let us consider in detail the form of the transmission coefficient
in the vicinity of of $E_m$.
Using the asymptotic expression (\ref{asympQ}) for $\widetilde Q_{ij}(E)$,
we obtain the following representation for $t_{21}(k)$
\begin{equation}
                \label{reson}
t_{21}(k)\simeq c_m\frac{k_m}{f_m+k_m},
\end{equation}
as $E \to E_m$.
Here $c_m$
is a normalization factor and
\begin{equation}
                \label{fm}
f_m=\frac{\det\alpha^{(m)}}{\gamma_m}
\end{equation}
is a complex number with
\begin{eqnarray}
\gamma_m=&&\alpha^{(m)}_{11}\left(\beta_{22}-
\frac{2\kappa_m\lambda^A_2}{(\kappa_m\lambda^C_2)^2+16}\right)+
\nonumber\\
&&\alpha^{(m)}_{22}\left(\beta_{11}-
\frac{2\kappa_m\lambda^A_1}{(\kappa_m\lambda^C_1)^2+16}\right)-
\nonumber\\
&&\alpha^{(m)}_{12}\beta_{21}-\alpha^{(m)}_{21}\beta_{12}.\nonumber
\end{eqnarray}

\begin{figure}[!h]
{\centering
\epsfclipon
\epsfxsize=84mm
\epsffile{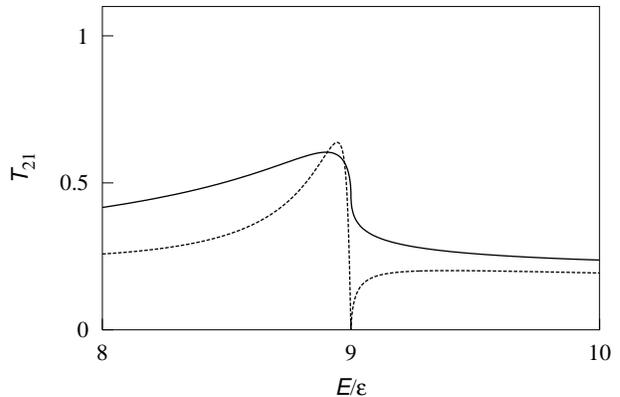}
\caption{\label{f-reson}
Transmission coefficient $T_{21}$
at $B=0$, $z_1-z_2=0.1 r $,
$\lambda^A_j=\lambda^B_j=\lambda^C_j=0.4  r $.
Solid line: $\varphi_1-\varphi_2=0$;
dash line: $\varphi_1-\varphi_2=0.08\pi$.}}
\end{figure}
One can see, that the behavior of transmission coefficient in the
vicinity of $E_m$ is determined by the value $f_m$.
Two curves for different positions of the contacts
corresponding to different $f_m$ are represented in Fig.~\ref{f-reson}.
If $f_m\to 0$ then
the transmission coefficient has a peak in a vicinity of
the zero $k=\kappa_p$.
The distance between the peak and the zero decreases with
decreasing  of $|f_m|$ while the peak value remains  finite.
The form of the graph in this region (dash line in Fig.~\ref{f-reson})
resembles the form of the Fano resonance,
but it is significant that the Fano curve
is smooth in contrast to function (\ref{reson}).
\begin{figure}[!h]
{\centering
\epsfclipon
\epsfxsize=84mm
\epsffile{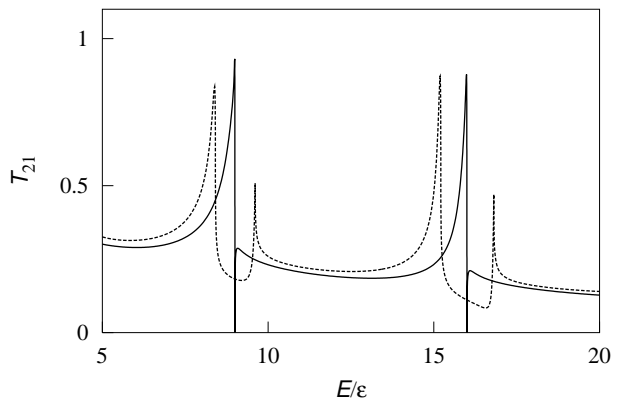}
\caption{\label{f-field}
Transmission coefficient $T_{21}$
at $\lambda^A_j=\lambda^B_j=\lambda^C_j=0.4  r $,
$z_1-z_2=0.05 r $,
$\varphi_1-\varphi_2=0.05\pi$.
Solid line: $B=0$;
dash line: $\Phi/\Phi_0=0.1$.}}
\end{figure}
\begin{figure}[!h]
{\centering
\epsfclipon
\epsfxsize=84mm
\epsffile{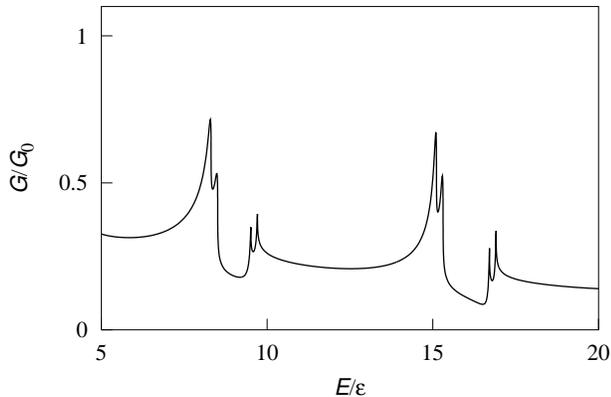}
\caption{\label{f-G}
Zero--temperature conductance $G$ of the system
as a function of the electron energy $E$
at $\Phi/\Phi_0=0.1$.
Other parameters are the same as in Fig.~\ref{f-field}
}}
\end{figure}
If $f_m=0$ then the peak and the zero
of transmission coincide and cancel each other
(solid line in Fig.~\ref{f-reson}).
We note that $f_m$ equals zero only if $\det\alpha^{(m)}=0$
as it follows from Eq.~(\ref{fm}).
Therefore the form of the transmission coefficient
in the vicinity of $E_m$ is determined
by the degree of degeneracy of $E_m$ and
by the symmetry of contact location.
In particular $\det\alpha^{(m)}=0$ for all
positions of contacts if the eigenvalue $E_m$ is non-degenerated
as it follows from Eq.~(\ref{alpha}).
Therefore the zeros do not appear
in a magnetic field with non-integer value of $2\Phi/\Phi_0$.
The magnetic field splits double degenerated energy levels $E_m$
and removes transmission zeros.
The dependence $T_{21}(E)$ for non-integer value of magnetic flux is
represented in Fig.~\ref{f-field}.
One can see that the peaks on the plot $T_{21}(E)$
transform into doublets.

The value $\det\alpha$ for integer $2\Phi/\Phi_0$ is given by
\begin{equation}
\label{det}
\det\alpha^{(m)}=-{(\pi  r )^{-2}}
\sin^2\left( (m+\Phi/\Phi_0)(\varphi_2-\varphi_1)\right).
\end{equation}
If $\sin\left( (m+\Phi/\Phi_0)(\varphi_2-\varphi_1)\right)=0$
then the value $f_m$ vanishes and the zero at the point $E_m$
disappears.
This phenomenon is similar to the collapse of the Fano resonance in the
transmission through a quantum sphere \cite{BGM02,GMP03}.
The disappearance of the zeros is associated with the symmetry
of the contact location.
We note that all zeros disappear if the
points ${\bf q}_1$ and ${\bf q}_2$ are placed on the same
generatrix ($\varphi_1=\varphi_2$) or on the
opposite generatrices of the cylinder
($|\varphi_1-\varphi_2|=\pi$).
It is significant that the positions of all zeros
are independent of the scattering lengths
$\lambda^A_j$, $\lambda^B_j$, and $\lambda^C_j$.

In the case of $f_m=0$
the transmission coefficient $T_{21}$
may be represented near $E_m$ in the form
$$
T_{21}(k)\simeq |c_m|^2 |1+g_m k_m|^2,
$$
where $c_m$ is a normalization factor and  $g_m$ is a complex number
depending on the position of contacts and scattering lengths.
The smoothness of the curve $T_{21}(E)$
at the point $E=E_m$ is determined by the number
$g_m$. Indeed, the
left-hand derivative is infinite in this point
if $\Im g_m \neq 0$ and the
right-hand derivative is infinite for $\Re g_m \neq 0$.
If the one-sided derivatives
are different then the transmission coefficient
has a kink at the point $E_m$.
The solid line in Fig.~\ref{f-reson}
corresponds to the case of infinite derivatives.

According to
Landauer--B\"uttiker formalism
the ballistic conductance
$G$ of the device at the zero temperature is determined
by transmission probabilities
$T_{21}^\uparrow$ and $T_{21}^\downarrow$
\begin{equation}
G=G_0(T_{21}^\uparrow+T_{21}^\downarrow),
\end{equation}
where $G_0=e^2/(2\pi\hbar)$ is the conductance quantum.
The conductance as a function of the electron energy
is shown in Fig.\ref{f-G}.
The dependence of the transmission coefficient
on electron spin orientation in the magnetic field
results in additional splitting of conductance
peaks (Fig.~\ref{f-G}) and partial spin polarization
of transmitted electrons.
It should be mentioned that
the spin splitting $g\mu_B B$ is
independent of magnetic quantum number $m$
while the splitting of the eigenvalues $E_m$ is
proportional to $m$ hence the peaks are not equidistant.
It is essential that the spin polarization can
be changed either by magnetic field or by electron energy.
The complete polarization is possible
for integer and half-integer values of
magnetic flux $\Phi/\Phi_0$.

\begin{figure}[!h]
{\centering
\epsfclipon
\epsfxsize=84mm
\epsffile{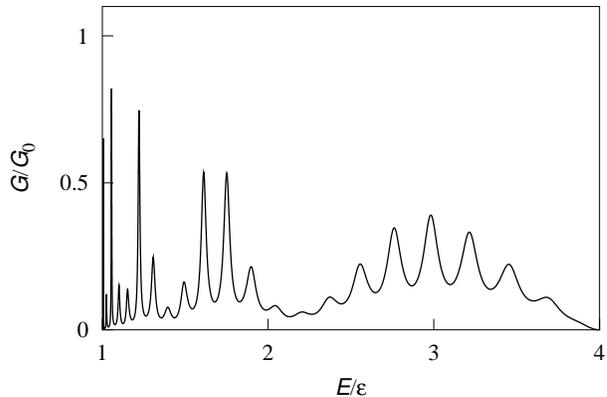}
\caption{\label{f-osc}
Zero--temperature conductance $G$ of the system
as a function of the electron energy
at $z_1-z_2=40 r $,
$\varphi_1=\varphi_2$,
$\lambda^A_j=\lambda^B_j=\lambda^C_j=0.4 r $,
and $B=0$.}}
\end{figure}

The conductance oscillates as a function of energy with
the period
\begin{equation}
\label{period}
\Delta E \simeq \frac {\pi}{L}\frac {dE}{dk_m}=\frac{\pi h^2 k_m}{4 m_{\rm c} L},
\end{equation}
if the longitudinal distance $L=|z_1-z_2|$
between the points ${\bf q}_1$ and ${\bf q}_2$
is much larger than the radius $r$
(see Fig.~\ref{f-osc}).
The oscillations are stipulated by the interference of
electron waves on the cylinder.
It should be noted that similar oscillations have been observed
in recent experiment \cite{KYT01} with carbon nanotubes.
The geometry of the experiment differs from ours,
in particular, contacts are not point-like.
But some results are valid for our system as well,
in particular, the estimation for the period of oscillations
cared out in Ref.~\onlinecite{KYT01} is in agreement with Eq.~(\ref{period}).

\section{Conclusion}

Electron transport in a nanodevice consisting of
a conductive cylinder with one-dimensional
wires connected to its lateral surface is considered.
The one-particle Hamiltonian of the system is obtained using
linear boundary conditions at the points of contact.
An explicit form for transmission and reflection coefficient
as a function of electron energy
is found solving the Schr\"odinger equation.
The general case of arbitrary number of wires and arbitrary
disposition of contacts is considered.
Two cases corresponding to a single wire and two wires attached to the
cylinder surface are studied in detail.
It is found that reflection peaks occur at
energies coinciding with the discrete part $E_m$
of the electron spectrum on the cylinder.
The form of reflection peaks is discussed.

A similar analysis of the two-wire case shows
that the transmission coefficient
equals zero at energies $E_m$.
We have found that asymmetric Fano-type resonances
appear in a vicinity of the zeros.
The zeros exist only if the number of magnetic flux quanta
through the cylinder is integer or half-integer.
They exist for all positions of contacts ${\bf q}_1$ and ${\bf q}_2$
except some specific points.
It is shown that the zero at the point $E_m$
disappears if the value $\det\alpha^{(m)}$
defined by Eq.~(\ref{det}) vanishes.
The behavior of the transmission coefficient in this case
resembles the collapse of the Fano resonances discussed
in earlier studies \cite{GMP03,KSJ99}.

The conductance of the device is investigated
using Landauer--Buttiker formalism.
The resonances in transmission coefficient lead to
appearance of conductance oscillations.
The magnetic field split conductance peaks and
cause spin polarization of transmitted electrons.
The complete spin polarization is possible for integer and
half-integer values of the magnetic flux.

The results of the paper may be useful for the study of electron
transport in single-wall carbon nanotubes and
rolled GaAs/AlGaAs heterostructures.
The experimental observation of the discussed effects
should become possible involving leads thin enough, like
the tip of the scanning tunnel microscope.
The geometry of the device in the case of one wire
resembles the geometry of experiments
on scanned probe microscopy of carbon nanotubes \cite{BFP00,YPR04}.
Experimental setup using two tips on the same nanotube
seems in principle feasible, although perhaps difficult to realize.
In the case of multi-mode leads the interference
of electron waves from different modes will most probably
result in additional transmission peaks and minima.
We stress that most of the obtained results
reflect the intrinsic properties of electron motion on the cylinder.
Therefore they are expected to remain valid qualitatively even
in the case of realistic non-one-dimensional wires.

\begin{acknowledgments}
This work is financially supported by
the Russian Foundation for Basic Research,
Grant No 05-02-16145.
\end{acknowledgments}

\end{document}